\documentclass{llncs}
\usepackage{epsf}

\newcommand{\almazero}{{\sf Alma-0}}
\newcommand{\alma}{{\sf Alma}}
\newcommand{\aaa}{{\sf AAA}}
 
\newcommand{\A}{\mbox{$\ \wedge\ $}}
\newcommand{\Or}{\mbox{$\ \vee\ $}}
\newcommand{\fa}{\mbox{$\forall$}}
\newcommand{\te}{\mbox{$\exists$}}
\newcommand{\ra}{\mbox{$\:\rightarrow\:$}}
\newcommand{\C}[1]{\mbox{$\{{#1}\}$}}           
\newcommand{\LL}{\mbox{$\ldots$}}
\newcommand{\NI}{\noindent}

\title{The Alma Project, or How First-Order Logic Can Help Us
      in Imperative Programming}

\author{Krzysztof R. Apt\inst{1,2} \and Andrea Schaerf\inst{3}}
\authorrunning{Krzysztof R. Apt and Andrea Schaerf}
\institute{CWI\\
P.O. Box 94079, 1090 GB Amsterdam, The Netherlands\\
\email{K.R.Apt@cwi.nl}
\and
Dept.\ of Mathematics, Computer Science, Physics \&
Astronomy 
University of Amsterdam, The Netherlands
\and
Dipartimento di Ingegneria Elettrica, Gestionale e Meccanica\\
Universit\`a di Udine\\
via delle Scienze 208, I-33100 Udine, Italy\\
\email{schaerf@uniud.it}
}
\begin{document}
\maketitle

\begin{abstract}
  The aim of the \alma{} project is the design of a strongly typed
  constraint programming language that combines the advantages of
  logic and imperative programming.
  
  The first stage of the project was the design and implementation of
  \almazero{}, a small programming language that provides a support for
  declarative programming within the imperative programming framework.
  It is obtained by extending a subset of Modula-2 by a small number
  of features inspired by the logic programming paradigm.
  
  In this paper we discuss the rationale for the design of
  \almazero{}, the benefits of the resulting hybrid programming
  framework, and the current work on adding constraint processing
  capabilities to the language. In particular, we discuss the role of
  the logical and customary variables, the interaction between the
  constraint store and the program, and the need for lists.

\end{abstract}

\section{Introduction}

\subsection{Background on Designing Programming Languages}

The design of programming languages is one of the most hotly debated
topics in computer science. Such debates are often pretty chaotic
because of the lack of universally approved criteria for evaluating
programming languages. In fact, the success or failure of a language
proposal often does not say much about the language itself but rather
about such accompanying factors as: the quality and portability of the
implementation, the possibility of linking the language with the
currently
reigning programming language standard (for instance, C), the existing
support within the industry, presence of an attractive development
environment, the availability on the most popular platforms, etc.

The presence of these factors often blurs the situation because in
evaluating a language proposal one often employs, usually implicitly,
an argument that the ``market'' will eventually pick up the best
product. Such a reasoning would be correct if the market forces in
computing were driven by the desire to improve the quality of
programming.  But from an economic point of view such
aspects as compatibility and universal availability are far more
important than quality.

Having this in mind we would like to put the above factors in a proper
perspective and instead concentrate on the criteria that have been used
in academia and which appeal directly to one of the primary purposes for
which a programming language is created, namely, to support an
implementation of the algorithms.  In what follows we concentrate on
the subject of ``general purpose'' programming languages, so the ones
that are supposed to be used for developing software, and for teaching
programming.

Ever since Algol-60 it became clear that such programming languages
should be ``high-level'' in that they should have a sufficiently rich
repertoire of control structures. Ever since C and Pascal it became
clear that such programming languages should also have a sufficiently
rich repertoire of data structures.

But even these seemingly obvious opinions are not universally accepted
as can be witnessed by the continuing debate between the followers of
imperative programming and of declarative programming. In fact, in
logic programming languages, such as Prolog, a support for just one
data type, the lists, is provided and the essence of declarative
programming as embodied in logic and functional programming lies in
not using assignment.

Another two criteria often advanced in the academia are that the
programming language should have a ``simple'' semantics and that the
programs should be ``easy'' to write, read and verify. What is
``simple'' and what is
``easy'' is in the eyes of the beholder, but both criteria can be used
to compare simplicity of various programming constructs and can be
used for example to argue against the {\tt goto} statement or 
pointers.

In this paper we argue that these last two criteria can be realized by
basing a programming language on first-order logic.  The point is that
first-order logic is a simple and elegant formalism with a clear
semantics. From all introduced formalisms (apart from the propositional
logic that is too simplistic for programming purposes) it is the one
that we understand best, both in terms of its syntax and its
semantics.  Consequently, its use should facilitate program
development, verification and understanding.

One could argue that logic programming has realized this approach to
computing as it is based on Horn clauses that are special types of
first-order formulas.  However, in logic programming in its original
setting computing (implicitly) takes place over the domain of terms.
This domain is not sufficient for programming purposes.
Therefore in Prolog, the most widely used logic programming language,
programs are augmented with some support for arithmetic. This
leads to a framework in which the logical basis is partly lost due to
the possibility of errors.  For instance, Prolog's assignment
statement {\tt X is t} yields a run-time error if {\tt t} is not a
ground arithmetic expression.

This and other deficiencies of Prolog led to the rise of constraint
logic programming languages that overcome some of Prolog's
shortcomings.  These programming languages depend in essential way on
some features as the presence of constraint solvers (for example a
package for linear programming) and constraint propagation.  So this
extension of logic programming goes beyond first-order logic.

It is also useful to reflect on other limitations of these two
formalisms. Both logic programming and constraint logic programming
languages rely heavily on recursion and the more elementary and easier
to understand concept of iteration is not available as a primitive.
Further, types are absent. They can be added to the logic
programming paradigm and in fact a number of successful proposals have
been made, see, e.g., \cite{Pfe92}. But to our knowledge
no successful proposal dealing with addition of types to
constraint logic programs is available.

Another, admittedly debatable, issue is assignment, shunned in logic
programming and constraint logic programming because its use destroys
the declarative interpretation of a program as a formula.  However, we
find that assignment {\em is\/} a useful construct. Some uses of it,
such as recording the initial value of a variable or counting the
number of bounded iterations, can be replaced by conceptually simpler
constructs but some other uses of it such as for counting or for
recording purposes are much less natural when simulated using logic
formulas.

\subsection{Design Decisions}
\label{subsec:design}

These considerations have led us to a design of a programming language
\almazero{}. 
The initial work on the design of this language was reported in
\cite{AS97}; the final description of the language, its
implementation and semantics is presented in \cite{ABPS98a}.

In a nutshell, \almazero{} has the following characteristics:

\begin{itemize}

\item 
it is an extension of a subset of Modula-2 that includes
  assignment, so it is a strongly typed imperative language;

\item to record the initial value of a variable the equality can be
used;

\item it supports so-called ``don't know'' nondeterminism by providing a
  possibility of a creation of choice points and automatic
  backtracking;

\item it provides two forms of bounded iterations.
\end{itemize}

The last two features allow us to dispense with many uses of recursion
that are in our opinion difficult to understand and to reason about.

As we shall see, the resulting language proposal makes programming in
an imperative style easier and it facilitates (possibly automated)
program verification. Additionally, for several algorithmic problems
the solutions offered by \almazero{} is substantially simpler than the
one offered by the logic programming paradigm.

The following simple example can help to understand what we mean by
saying that \almazero{} is based on first-order logic and that some
\almazero{} programs are simpler than their imperative and logic
programming counterparts.
 
Consider the procedure that tests whether an array {\tt a[1..n]} is
ordered.  The customary way to write it in Modula-2 is:

\begin{small}
\begin{verbatim}
i:= 1;
ordered := TRUE;
WHILE i < n AND ordered DO
  ordered := ordered AND (a[i] <= a[i+1]);
  i := i+1
END;
\end{verbatim}
\end{small}

In \almazero{} we can just write:

\begin{small}
\begin{verbatim}
ordered := FOR i:= 1 TO n-1 DO a[i] <= a[i+1] END
\end{verbatim}
\end{small}

This is much simpler and as efficient.  In fact, this use of the {\tt
FOR} statement corresponds to the bounded universal quantification and
the above one line program equals the problem specification.

In the logic programming framework there are no arrays. But the related
problem  of finding whether a list {\tt L} is ordered
is solved by the following program which is certainly more involved
than the above one line of \almazero{} code:

\begin{small}
\begin{verbatim}
ordered([]).
ordered([X]).
ordered([X, Y | Xs]) :- X =< Y, ordered([Y| Xs]).
\end{verbatim}
\end{small}

\subsection{Towards an Imperative Constraint Programming Language}

In \almazero{} each variable is originally {\em uninitialized} and
needs to be initialized before being used. Otherwise a run-time error
arises. The use of uninitialized variables makes it possible to use a
single program for a number of purposes, such as computing a solution,
completing a partial solution, and testing a candidate solution.  On
the other hand, it also provides a limitation on the resulting
programming style as several first-order formulas, when translated to
\almazero{} syntax, yield programs that terminate in a run-time error.

With the addition of constraints this complication would be overcome.
The idea is that the constraints encountered during the program
execution are moved to the constraint store and evaluated later, when
more information is available.  Then the above restriction that each
variable has to be initialized before being used can be lifted, at
least for the variables that are manipulated by means of constraints.
Additionally, more programs can be written in a declarative way.  In
fact, as we shall see, an addition of constraints to \almazero{} leads
to a very natural style of programming in which the constraint
generation part of the program is often almost identical to the
problem specification.

Constraint programming in a nutshell consists of generating
constraints (requirements) and solving them by general and domain
specific methods.  This approach to programming was successfully
realized in a number of programming languages, notably constraint
logic programming languages.

Up to now, the most successful approach to imperative constraint
programming is the object-oriented approach taken by ILOG Solver (see
\cite{PL95}, \cite{ILOG98}).  In this system constraints
and variables are treated as objects and are defined within a C++
class library. Thanks to the class encapsulation mechanism and the
operator overloading capability of C++, the user can see constraints
almost as if they were a part of the language.  A similar approach was
independently taken in the {\em NeMo+\/} programming environment of
\cite{STU97}.

In our approach constraints are integrated into the imperative
programming paradigm, as ``first class citizens'' of the language.
The interaction between the constraint store and the program becomes
then more transparent and conceptually simpler
and the resulting constraint programs are in our opinion more natural
than their counterparts written in the constraint logic programming
style or in the imperative languages augmented with constraint
libraries.

The reason for this in the case of constraint logic programming is
that many uses of recursion and lists can be replaced by the more
basic concepts of bounded iteration and arrays. In the case of the
imperative languages with constraint libraries, due to the absence of
non-determinism in the language, failure situations (arising due to
inconsistent constraints) must be dealt with explicitly by the
programmer, whereas in \almazero{} they are managed implicitly by the
backtracking mechanism.

When adding constraints to a strongly typed imperative programming
language one needs to resolve a number of issues.  First, constraints
employ variables in the mathematical sense of the word (so {\em
  unknowns\/}) while the imperative programming paradigm is based on
the computer science concept of a variable, so a {\em known\/}, but
varying entity.  We wish to separate between these two uses of
variables because we want to manipulate unknowns only by means of
constraints imposed on them.  This precludes the modelling of unknowns
by means of uninitialized variables since the latter can be modified by
means of an assignment.

Second, one needs to integrate the constraints in such a way that
various features of the underlying language such as use of local and
global declarations and of various parameter passing mechanisms
retain their coherence.

Additionally, one has to maintain the strong typing discipline according
to which each variable has a type associated with it in such a way
that throughout the program execution only values from its type can be
assigned to the variable.  Finally, one needs to provide an adequate
support for search, one of the main aspects of constraint programming.

So the situation is quite different than in the case of the logic
programming framework.  Namely, the logic programming paradigm is
based on the notion of a variable in the mathematical sense (usually
called in this context a {\em logical\/} variable).  This greatly
facilitates the addition of constraints and partly explains why the
integration of constraints into logic programming such as in the case
of CHIP (see \cite{vanhentenryck-constraint-using}), Prolog III
(see \cite{Col90}) and CLP(${\cal R}$) (see
\cite{JAFFAR92C}), to name just three examples, has been so
smooth and elegant.  Further, logic programming languages provide
support for automatic backtracking.

However, as already mentioned, in constraint logic programming
languages types are not available. Moreover, there is a very limited
support
for scoping and only one parameter mechanism is available. 

Let us return now to \almazero{}. The language already provides a
support for search by means of automatic backtracking. This support is
further enhanced in our proposal by providing a built-in constraint
propagation.  In \cite{ABPS98a} we stated that our language
proposal should be viewed as ``an instance of a {\em generic method\/}
for extending (essentially) any imperative programming language with
facilities that encourage declarative programming.''  That is why we
think that the proposal here discussed should be viewed not only as a
suggestion how to integrate constraints into \almazero{}, but more
generally how to integrate constraints into any strongly typed
imperative language.  In fact, \almazero{} can be viewed as an
intermediate stage in such an integration.

The remainder of the paper is organized as follows. In
Section~\ref{sec:alma-0} we
summarize the new features of \almazero{} and in
Section~\ref{sec:programming} we
illustrate the resulting programming style by two examples. Then, in
Section~\ref{sec:adding-constraints} we discuss the basics of our
proposal for adding constraints to \almazero{} and in
Section~\ref{sec:procedures} we explain how constraints interact with
procedures. In turn in Section~\ref{sec:extensions} we discuss language
extensions for expressing complex constraints and for facilitating
search in presence of constraints. Finally, in
Section~\ref{sec:related-work} we discuss related work and in
Section~\ref{sec:conclusions} we draw some conclusions and discuss
the future work.

\section{A Short Overview of \almazero{}}
\label{sec:alma-0}

\almazero{} is an extension of a subset of Modula-2 by nine new
features inspired by the logic programming paradigm. We briefly recall
most of them here and refer to \cite{ABPS98a} for a detailed
presentation.

\begin{itemize}
\item Boolean expressions can be used as statements and vice versa.
  This feature of \almazero{} is illustrated by the above one line
  program of Subsection \ref{subsec:design}.

  A boolean expression that is used as a statement and evaluates to
  {\tt FALSE} is identified with a {\em failure}.

\item {\em Choice points} can be created by the non-deterministic
  statements {\tt ORELSE} and {\tt SOME}. The former is a dual of the 
  statement composition and the latter is a dual of the {\tt FOR}
  statement.  Upon failure the control returns to the most recent
  choice point, possibly within a procedure body, and the computation
resumes with the next branch in the state in which the previous branch
was entered.
  
\item The created choice points can be erased or iterated over by
means of the {\tt COMMIT} and {\tt FORALL} statements. {\tt COMMIT S
END} removes the choice points created during the first successful
execution of {\tt S}.  {\tt FORALL S DO T END} iterates over all
choice points created by {\tt S}. Each time {\tt S} succeeds, {\tt T}
is executed.

\item The notion of {\em initialized} variable is introduced: A variable
is
  uninitialized until the first time a value is assigned to it; from
that point
  on, it is initialized. The {\tt KNOWN} relation tests whether a
variable of a
  simple type is initialized.
\item The equality test is generalized to an assignment statement in
case one
  side is an uninitialized variable and the other side an expression
with known
  value.

\item In \almazero{} three types of parameter mechanisms are allowed:
  call by value, call by variable and {\em call by mixed form}. The
  first two are those of Pascal and Modula-2; the third one is an
  amalgamation of the first two (see \cite{ABPS98a}).  Parameters
  passed by mixed form can be used both for testing and for computing.
\end{itemize}

Let us summarize these features of \almazero{} by clarifying which of
them
are based on first-order logic.

In the logical reading of the programming language constructs
the program composition {\tt S; T} is viewed as the conjunction
${\tt S \A {\tt T}}$.
A dual of ``;'', the {\tt EITHER S ORELSE T END} statement,
corresponds then to the disjunction ${\tt S \Or {\tt T}}$.

Further, the {\tt FOR i:= s TO t DO S END} statement is viewed as
the bounded universal quantification, ${\tt \fa i \in [s..t] \:S}$,
and its dual, the {\tt SOME i:= s TO t DO S END} statement is viewed as
the bounded existential quantification, ${\tt \te i \in [s..t] \: S}$.

In turn, the {\tt FORALL S DO T END} statement can be viewed as 
the restricted quantification ${\tt \fa \bar{x} (S \ra T)}$, where
${\tt \bar{x}}$ are all the variables of {\tt S}.

Because the boolean expressions are identified with the statements, we
can apply the negation connective, {\tt NOT}, to the statements.
Finally, the equality can be interpreted both as a test and as an
one-time assignment, depending on whether the variable in question is
initialized or not.

\section{Programming in \almazero{}}
\label{sec:programming}

To illustrate the above features of \almazero{} and the resulting
programming style we now consider two examples.

\subsection{The Frequency Assignment Problem}

The first problem we discuss is a combinatorial problem from
telecommunication.

\begin{problem} {\it Frequency Assignment\/}
(\cite{Hal80}). \label{problem-frequency}
  Given is a set of $n$ {\em cells}, $C := \{c_1,$ $c_2, \dots, c_n\}$
  and a set of $m$ frequencies (or channels) $F := \{f_1, f_2, \dots,
  f_m\}$.  An {\em assignment\/} is a function which associates with
each cell
  $c_i$ a frequency $x_i \in F$. The problem consists in finding an
  assignment that satisfies the following constraints.

\begin{description}
  
\item[Separations:] Given $h$ and $k$ we call the value
  $d(f_h,f_k)=~\!\mid\!h-k\!\mid$ the {\em distance\/} between two
  channels $f_h$ and $f_k$. (The assumption is that consecutive
frequencies
  lie one unit apart.)  Given is an $n \times n$ non-negative integer
  symmetric matrix $S$, called a {\em separation matrix}, such that
  each $s_{ij}$ represents the minimum distance between the
  frequencies assigned to the cells $c_i$ and $c_j$. That is, for all
  $i \in [1..n]$ and $j \in [1..n]$ it holds that $d(x_{i},x_{j})
  \geq s_{ij}$.

\item[Illegal channels:] Given is an $n \times m$ boolean matrix $F$
  such that if $F_{ij} = true$, then the frequency $f_j$ cannot be
  assigned to the cell $i$, i.e., $x_i \neq f_j$.
\end{description}
\end{problem}

Separation constraints prevent interference between cells which are
located geographically close and which broadcast in each other's area of
service.  Illegal channels account for channels reserved for external
uses (e.g., for military bases).

The \almazero{} solution to this problem does not use an assignment
and has a dual interpretation as a formula.  We tested this program on
various data.  We assume here for simplicity that each $c_i$ equals
$i$ and each $f_i$ equals $i$, so $C = \C{1,\LL,n}$ and $F =
\C{1,\LL,m}$.

\begin{small}
\begin{verbatim}
MODULE FrequencyAssignment;
CONST N = 30;  (* number of cells *)
      M = 27;  (* number of frequencies *)

TYPE SeparationMatrix = ARRAY [1..N],[1..N] OF INTEGER;
     IllegalFrequencies = ARRAY [1..N],[1..M] OF BOOLEAN;
     Assignment = ARRAY [1..N] OF [1..M];  (* solution vector *)

VAR S: SeparationMatrix; 
    F: IllegalFrequencies;
    A: Assignment;
    noSol: INTEGER;

PROCEDURE AssignFrequencies(S: SeparationMatrix; F: IllegalFrequencies; 
                            VAR A: Assignment);
VAR i, j, k: INTEGER;
BEGIN
  FOR i := 1 TO N DO
    SOME j := 1 TO M DO (* j is a candidate frequency for cell i *)
      NOT F[i,j];
      FOR k := 1 TO i-1 DO
        abs(A[k] - j) >= S[k,i]
      END;
      A[i] = j
    END
  END
END AssignFrequencies;

BEGIN
  InitializeData(S,F);
  AssignFrequencies(S,F,A);
  PrintSolution(A)
END FrequencyAssignment.
\end{verbatim}
\end{small}

The simple code of the procedures {\tt InitializeData} and {\tt
  PrintSolution} is omitted.  
The generalized equality {\tt A[i] = j} serves here as an assignment
and the {\tt SOME} statement takes care of automatic backtracking
in the search for the right frequency {\tt j}.

In the second part of the paper we shall
discuss an alternative solution to this problem using constraints.

\subsection{Job Shop Scheduling}

The second problem we discuss is a classical scheduling problem,
namely the {\em job shop scheduling\/} problem. We refer to
\cite[page 242]{GJ79} for its precise description.  Roughly
speaking, the problem consists of scheduling over time a set of jobs,
each consisting of a set of consecutive tasks, on a set of processors.

The input data is represented by an array of jobs, each element of
which is a record that stores the number of the tasks and the array of
tasks.
In turn, each task is represented by the machine it uses and by its
length.  The output is delivered as an integer matrix that (like a
so-called Gantt chart) for each time point $k$ and each processor $p$
stores the job number that $p$ is serving at the time point $k$.

The constraint that each processor can perform only one job at a time
is enforced by using generalized equality on the elements of the
output matrix. More precisely, whenever job $i$ requires processor $j$
for a given time window $[d_1,d_2]$, the program attempts for some $k$
to initialize the elements of the matrix $(j,k + d_1), (j,k +
d_1+1),\dots,(j,k + d_2)$ to the value $i$. If this initialization
succeeds, the program continues with the next task. Otherwise some
element in this segment is already initialized, i.e., in this segment
processor $j$ is already used by another job.  In this case the
execution fails and through backtracking the next value for $k$ is
chosen.

The constraint that the tasks of the same job must be executed in the
provided order and cannot overlap in time is enforced by the use of
the variable {\tt min\_start\_time} which, for each job, initially
equals 1 and then is set to the end time of the last considered task
of the job. To perform this update we exploit the fact that when the
{\tt SOME} statement is exited its index variable {\tt k} equals the
smallest value in the considered range for which the computation does
not fail (as explained in \cite{ABPS98a}).

We provide here the procedure that performs the scheduling. For the
sake of brevity the rest of the program is omitted.

\begin{small}
\begin{verbatim}
TYPE
   TaskType      = RECORD
                      machine : INTEGER;
                      length  : INTEGER;
                   END;
   JobType       = RECORD
                      tasks : INTEGER;
                      task  : ARRAY [1..MAX_TASKS] OF TaskType
                   END;
   JobVectorType = ARRAY [1..MAX_JOBS] OF JobType;
   GanttType     = ARRAY [1..MAX_MACHINES],[1..MAX_DEADLINE] OF INTEGER;

PROCEDURE JobShopScheduling(VAR job: JobVectorType; deadline:INTEGER; 
                            jobs :INTEGER; VAR gantt: GanttType);
VAR
   i, j, k, h     : INTEGER;
   min_start_time : INTEGER;   
BEGIN             
  FOR i := 1 TO jobs DO
    min_start_time := 1;
    FOR j := 1 TO job[i].tasks DO
      SOME k := min_start_time TO deadline - job[i].task[j].length + 1
DO
        (* job i engages the processor needed for task j from time k to
           k + (length of task j) - 1.
           If the processor is already engaged, the program backtracks.
*)
        FOR h := k TO k + job[i].task[j].length - 1 DO
          gantt[job[i].task[j].processor,h] = i;
        END  
      END;
      min_start_time := k + job[i].task[j].length;
        (* set the minimum start time for the next task 
           to the end of the current task *)
    END;  
  END
END JobShopScheduling;
\end{verbatim}
\end{small}

In this program the ``don't know'' nondeterminism provided by the use
of the {\tt SOME} statement is combined with the use of assignment.

Furthermore, as already mentioned, for each value of {\tt i} and {\tt j}
the equality {\tt gantt[job[i].task[j].processor,h] = i} acts both as
an assignment and as a test. 

The array {\tt gantt} should be uninitialized when the procedure is
called. At the end of the execution the variable {\tt gantt} contains
the first feasible schedule it finds.

Preinitialized values can be used to enforce some preassignments of
jobs to processors, or to impose a constraint that a processor is not
available during some periods of time.  For example, if processor $2$
is not available at time $5$, we just use the assignment {\tt
gantt[2,5] := 0} (where $0$ is a dummy value) before invoking the
procedure {\tt JobShopSchedule}.

As an example, suppose we have 3 jobs, 3 processors ($p_1$, $p_2$, and
$p_3$), the deadline is 20, and the jobs are composed as follows:

\begin{small}
\begin{center}
\begin{tabular}[tbh]{|l|c|cc|cc|cc|cc|}\hline
      &        & \multicolumn{2}{c|}{task 1} & \multicolumn{2}{c|}{task
2} 
               & \multicolumn{2}{c|}{task 3} & \multicolumn{2}{c|}{task
4}\\
job   & tasks  & proc & len  & proc & len & proc & len & proc & len \\
\hline
1     & 4      & $p_1$ & 5 &  $p_2$ & 5 &  $p_3$ & 5 &  $p_2$ & 3\\
2     & 3      & $p_2$ & 6 &  $p_1$ & 6 &  $p_3$ & 4 &        &  \\
3     & 4      & $p_3$ & 6 &  $p_2$ & 4 &  $p_1$ & 4 &  $p_2$ & 1\\
\hline
\end{tabular}
\end{center}
\end{small}

The first solution (out of the existing 48) for the array {\tt gantt}
that the program finds is the following one, where the symbol {\tt
  '-'} means that the value is uninitialized, i.e., the processor is
idle in the corresponding time point.

\begin{small}
\begin{verbatim}
    1  1  1  1  1  -  2  2  2  2  2  2  -  -  -  3  3  3  3  - 
    2  2  2  2  2  2  1  1  1  1  1  3  3  3  3  -  1  1  1  3 
    3  3  3  3  3  3  -  -  -  -  -  1  1  1  1  1  2  2  2  2 
\end{verbatim}
\end{small}

\vspace{.5\baselineskip}

For some applications, it is necessary to make the schedule as short
as possible. To this aim, we can use the following program fragment.

\begin{small}
\begin{verbatim}
COMMIT
  SOME deadline := 1 TO max_deadline DO
    JobShopScheduling(JobVector,deadline,jobs,Gantt)
  END
END
\end{verbatim}
\end{small}

It computes the shortest schedule by {\em guessing}, in ascending order,
the first deadline that can be met by a feasible assignment.
The use of the {\tt COMMIT} statement ensures that once a solution
is found, the alternatives, with larger {\tt deadline} values,
are discarded.

\section{Introducing Constraints}\label{sec:adding-constraints}

In what follows we discuss a proposal for adding constraints to
\almazero{}.


This Section is organized as follows.
In Subsection~\ref{subsec:adding-unknowns} we discuss the addition of
constrained types and unknowns to the language and in
Subsections~\ref{subsec:constraint-store}
and~\ref{subsec:store-vs-program} we define the constraint store and
illustrate its interaction with the program execution.

To illustrate how the proposed addition of constraints to \almazero{}
provides a better support for declarative programming we illustrate in
Subsection~\ref{subsec:examples} their use by means of three example
programs.

To simplify our considerations we ignore in this section the presence of
procedures. In particular, we assume for a while that all declarations
are at
one level.

\subsection{Adding Constrained Types, Unknowns and Constraints} 
\label{subsec:adding-unknowns}

We start by adding a new kind of variables of simple types, called
unknowns.  This is done by using the qualifier {\tt CONSTRAINED} in
declarations of {\em simple types}, that is {\tt INTEGER}, {\tt
  BOOLEAN}, {\tt REAL}, enumeration and subrange types.

\begin{definition}~\\[-6mm]
\begin{itemize}
\item A type qualified with the keyword {\tt CONSTRAINED} is called a
{\em
   constrained type}.
\item A variable whose type is a constrained type is called an {\em
unknown}.
\end{itemize}
\end{definition}

We shall see in Section \ref{sec:procedures} that this way of
defining unknowns simplifies the treatment of parameter passing in
presence of unknowns.  From now on we distinguish between variables
and unknowns.  In the discussion below we assume the following
declarations.

\small \begin{verbatim}
CONST N = 8;
TYPE Board = ARRAY [1..N] OF CONSTRAINED [1..N];
     Colour = (blue, green, red, yellow);
     Info = RECORD 
       co: Colour; 
       No: CONSTRAINED INTEGER;
     END;
VAR i, j: INTEGER;
    a: ARRAY [1..N] of INTEGER;
    C: CONSTRAINED [1..N];
    X, Y: Board;
    Z: Info;
\end{verbatim} 
\normalsize

So {\tt a}, {\tt i} and {\tt j} are variables while {\tt C} is an
unknown.  In turn, {\tt X} and {\tt Y} are arrays of unknowns and {\tt
  Z} is a record the first component of which is a variable and the
second an unknown.

Because of the syntax of \almazero{}, boolean expressions can appear
both in the position of a statement and inside a condition.  

\begin{definition} \label{def:constraint}
  A {\em constraint} is a boolean expression that involves some
  unknowns. 

We postulate that the unknowns can appear only within constraints
or within the right hand side of an assignment.
\end{definition}

The values of unknowns are determined only by means of constraints
that are placed on them.  In particular, by the just introduced
syntactic restriction, one cannot use assignment to assign a value to
an unknown. So in presence of the above declarations the statements
{\tt X[1] := 0} and {\tt C := 1} are illegal.  In contrast, the
constraints {\tt X[1] = 0} and {\tt C = 1} are legal.
Further, the assignments {\tt i := X[1] + X[2]} and {\tt i := Y[X[2]]}
are
also legal statements. 

Initially each unknown has an
{\em undetermined\/} value that belongs to the domain associated with
the
type. By placing constraints on an unknown its domain can {\em
shrink}.  The unknown continues to have an undetermined value until
the domain gets reduced to a singleton.

If the program control reaches an occurrence of an unknown outside of
a constraint, 
so within the right hand side of an assignment, this unknown is
evaluated. 
If its value is at this moment
undetermined, this evaluation yields a run-time error. If the value is
{\em determined\/} (that is, the domain is a singleton), then it is
substituted for the occurrence of the unknown. So the occurrences of
an unknown outside of a constraint are treated as usual variables.

Note that during the program execution the domain of an unknown {\em
  monotonically\/} decreases with respect to the subset ordering.
This is in stark contrast with the case of variables. Initially, the
value of a variable of a simple type is not known but after the first
assignment to it its value is determined though can {\em
  non-monotonically} change to any other value from its type.

Intuitively, a program is viewed as an ``engine''
that generates constraints. These constraints are gradually
solved by means of the constraint solving process that we shall explain
now.


\subsection{Adding the Constraint Store} \label{subsec:constraint-store}

We now introduce the central notion of a {\em constraint store}.  This
is done in a similar way as in the constraint logic programming
systems, though we need to take into account here the presence of
variables and constants.

\begin{definition}
  We call a constraint $C$ {\em evaluated\/} if each constant that
occurs
  in it is replaced by its value and each variable (not unknown) that
  occurs in it is replaced by its current value. If some variable that
  occurs in $C$ is uninitialized, we say that the evaluation of $C$
  {\em yields an error}.  Otherwise we call the resulting boolean
  expression the {\em evaluated form\/} of $C$.  
\end{definition}

So no variables occur in the evaluated form of a constraint.  For
technical reasons we also consider a {\em false constraint}, denoted
by $\bot$, that can be generated only by a constraint solver to
indicate contradiction.

\begin{definition} 
  A {\em constraint store\/}, in short a {\em store}, is a set of
  evaluated forms of constraints.  We say that an unknown is {\em
    present in the store\/} if it occurs in a constraint that belongs
  to the store. 

We call a store {\em failed} if $\bot$ is present in
  it or if the domain of one of the unknowns present in it is empty.
  By a {\em solution\/} to the store we mean an assignment of values
  from the current domains to all unknowns present in it.

Further, we say that a constraint is {\em solved\/} if its evaluated
form is satisfied by all combinations of values from the current
domains of its unknowns. 
\end{definition}

For example, in the program fragment

\small \begin{verbatim}
   i := 1;
   j := 2;
   X[i] <= j;
   Y[X[i+2]] <> Y[N];
\end{verbatim} 
\normalsize

\NI we have two constraints, {\tt X[i] <= j} and {\tt Y[X[i+2]] <>
  Y[N]}. Here {\tt X[1] <= 2} is the evaluated form of the first one,
while {\tt Y[X[3]] <> Y[8]} is the evaluated form of the second one.
If we deleted the assignment {\tt i := 1} the evaluations of both
constraints would yield an error.

The notion of a failed store is a computationally tractable
approximation of
that of an inconsistent store, i.e., a store that has no solutions.
Indeed, a
failed store is inconsistent but an inconsistent store does not have to
be
failed: just consider {\tt X[1] = X[2], X[1] $<$ X[2]}.

\subsection{Interaction between the Program and the Constraint Store}
\label{subsec:store-vs-program}

The program interacts with the store in the following two ways:

\begin{itemize}

\item By adding to it the evaluated forms of the encountered
constraints.
  If the evaluation of such a constraint yields an error, a run-time
  error arises.

\item By generating possible values for unknowns that are present in
  the store by means of some built-in primitives to be introduced in
  Subsection~\ref{subsec:indomain-built-in}.

\end{itemize}

The store is equipped with a number of procedures called {\em
  constraint solvers}.  Their form depends on the applications. One or
more of them can become activated upon addition of (an evaluated form
of) a constraint to the store. An activation of constraint solvers, in
the sequel called {\em constraint solving}, can reduce the domains of
the unknowns, determine the values of some unknowns by reducing the
corresponding domains to singletons, delete some constraints that are
solved, or discover that the store is failed, either by generating the
false
constraint $\bot$ or by reducing the domain of an unknown to the empty
set.

We assume that constraint solving is a further unspecified
process that depending of application may be some form 
of constraint propagation or a decision procedure. We 
require that the result of constraint solving
maintains equivalence, which means that the set of all solutions to
the store does not change by applying to it constraint
solvers.


The store interacts with the program as follows.

\begin{definition} 
Upon addition of a constraint to the store, constraint solving
    takes place.
\begin{itemize}
  \item If as a result of the constraint solving the store remains
    non-failed, the control returns to the program and the execution
    proceeds in the usual way.
  \item Otherwise the store becomes failed and a {\em failure\/}
    arises. This means that the control returns to the last choice
    point created in the program.  Upon backtracking all the
    constraints added after the last choice point are retracted and
    the values of the variables and the domains of the unknowns are
    restored to their values at the moment that the last choice point
was
    created. 
  \end{itemize}
\end{definition}

This means that we extend the notion of failure, originally introduced
in Section \ref{sec:alma-0}, to deal with the presence of the store.

Note that constraints are interpreted in the same way
independently of the fact whether they appear as a statement or inside a
condition. For example, the following program fragment

\begin{verbatim}
  IF X[1] > 0 THEN S ELSE T END
\end{verbatim}
is executed as follows: The constraint {\tt X[1] > 0} is added to the
store. If the store does not fail {\tt S} is executed, otherwise
{\tt T} is executed. So we do not check whether {\tt X[1] > 0} is {\em
  entailed\/} by the store and execute {\tt S} or {\tt T} accordingly,
as one might intuitively expect.  This means that constraints are
always interpreted as so-called {\em tell\/} operations in the store,
and never as so-called {\em ask\/} operations, which check for
entailment (see Section~\ref{sec:conclusions} for a discussion on this
point).

\subsection{Examples}
\label{subsec:examples}

To illustrate use of the introduced concepts we now consider three
examples. We begin with the following classical problem.

\begin{problem} {\it Eight Queens.\/}
Place 8 queens on the chess board so that they do not attack each
other.
\label{problem-queens}
\end{problem}

We present here a solution that uses constraints. We only write the
part of the program that generates constraints. The code that actually
solves the generated constraints would make use of the built-in {\tt
  INDOMAIN} as explained in Subsection~\ref{subsec:indomain-built-in}.

\small \begin{verbatim}
CONST N = 8;
TYPE Board = ARRAY [1..N] OF CONSTRAINED [1..N]; 
VAR i, j: [1..N];
    X: Board;

BEGIN
  FOR i := 1 TO N-1 DO 
    FOR j := i+1 TO N DO 
      X[i] <> X[j];
      X[i] <> X[j]+j-i;
      X[i] <> X[j]+i-j     
    END
  END
END;

\end{verbatim} 
\normalsize

Each generated constraint is thus of the form {\tt X[i] <>
  X[j]} or {\tt X[i] <> X[j] + k} for some values {\tt i,j $\in$
  [1..N]} such that ${\tt i < j}$ and {\tt k} being either the value
of {\tt j-i} or of {\tt i-j}.

Note that the above program text coincides with the problem formulation.

Next, consider the following problem that deals
with the equations arising when studying the flow of heat.

\begin{problem} {\it Laplace Equations}. 
Given is a two dimensional grid with given values for all the
exterior points.  The value of each interior points equals
the average of the values of its four neighbours.
Compute the value of all interior points.
\label{problem-laplace}
\end{problem}

The solution using constraints again truly coincides with the problem
specification. It is conceptually much simpler than the solution
based on constraint logic programming and given in \cite{JL87}.

\small \begin{verbatim}
TYPE Board = ARRAY [1..M], [1..N] OF CONSTRAINED REAL; 
VAR i:[1..M]; 
    j:[1..N]; 
    X: Board;

BEGIN
  FOR i := 2 TO M-1 DO 
    FOR j := 2 TO N-1 DO 
      X[i,j] = (X[i+1,j] + X[i-1,j] + X[i,j+1] + X[i,j-1])/4
    END
  END
END;
\end{verbatim} 
\normalsize

We assume here that the constraint solver that deals with linear
equations over reals is sufficiently powerful to solve the generated
equations.

Finally, we present a solution to the {\em Frequency Assignment} problem
(Problem~\ref{problem-frequency}) that uses constraints.  Again, we
only write the part of the program that generates constraints. We
assume here that the variables {\tt S} and {\tt F} are properly
initialized.

\small \begin{verbatim}
TYPE SeparationMatrix = ARRAY [1..N],[1..N] OF INTEGER;
     IllegalFrequencies = ARRAY [1..N],[1..M] OF BOOLEAN;
     Assignment = ARRAY [1..N] OF CONSTRAINED [1..M];
VAR S: SeparationMatrix;
    F: IllegalFrequencies;
    X: Assignment;
    i, j: INTEGER;

BEGIN
  FOR i := 1 TO N DO
    FOR j := 1 TO M DO
      IF F[i,j] THEN X[i] <> j END
    END
  END;
  FOR i := 1 TO N DO
    FOR j := 1 TO i-1 DO
      EITHER X[i] - X[j] >= S[i,j]
      ORELSE X[j] - X[i] >= S[i,j]
      END
    END
  END
END;
\end{verbatim} 
\normalsize

The use of the {\tt ORELSE} statement creates here choice points to
which the control can return if in the part of the program that deals
with constraints solving a failed store is produced.

Alternatively, one could use here a disjunction and replace the {\tt
ORELSE}
statement by
\begin{center} 
    {\tt (X[i] - X[j] >= S[j,i]) OR (X[j] - X[i] >= S[j,i])}.
\end{center}
In this case no choice points are created but the problem of solving
(disjunctive) constraints is now ``relegated'' to the store.

The latter solution is preferred if the constraint solver in use
is able to perform some form of preprocessing on disjunctive
constraints, such as the {\em constructive
  disjunction} of \cite{hentenryck-design-95}. On the
other hand, the former solution allows the programmer to retain
control upon the choice generated by the system. For example, she/he
can associate different actions to the two branches of the {\tt ORELSE}
statement.  

It is important to realize that the integration of constraints to
\almazero{} as outlined in this section is possible only because the
unknowns are initially uninitialized.

\section{Constraints and Procedures}\label{sec:procedures}

So far we explained how the program interacts with the store in
absence of procedures. In \almazero{} one level (i.e., not nested)
procedures are allowed.  In presence of procedures we need to explain
a number of issues.

First, to keep matters simple, we disallow local unknowns.
This means that the constrained types can be only introduced
at the outer level. However, unknowns can be used within the
procedure bodies provided the restrictions introduced in
Definition \ref{def:constraint} are respected.


Next, we need to explain how unknowns can be passed as parameters.
Formal parameters of constrained types are
considered as unknowns. This means that in the procedure body
such formal parameters can occur only within the constraints or 
within the right hand side of an assignment.

We discuss first call by variable. An unknown (or a compound variable
containing an unknown) passed as an actual variable parameter is
handled in the same way as the customary variables, by means of the
usual reference mechanism. 

For example consider the following problem.

\begin{problem} 
  Given is an array which assigns to each pixel on an ${\tt M} \times
{\tt N}$ board a
  colour.  A {\em region\/} is a maximal set of adjacent pixels that
have the
  same colour.  Determine the number of regions.
\end{problem}

\begin{figure}[htbp]
\epsfxsize 3.5cm
\centerline{\epsfbox{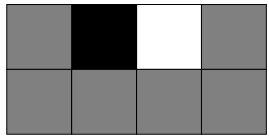}}  
\caption{Constraint Solving and Pixels}
\label{fig:pixels}
\end{figure}

To solve it we represent each pixel as a record, one field
of which holds the colour of the pixel and the other is an unknown
integer.  Then we assign to each pixel a number in such a way that
pixels in the same region get the same number.  These assignments are
performed by means of constraint solving.  For instance, in the
case of Figure \ref{fig:pixels} the constraint solving takes care
that the value 1 is assigned to all but two pixels once it is assigned
to the leftmost uppermost pixel.

To achieve this effect in the program below we assume that the
constraint solving process is able to reduce the domain of {\tt y} to
\C{{\tt a}} given the constraint {\tt x = y} and the fact that the
domain of {\tt x} equals \C{{\tt a}}. The program uses both
constraints and an assignment. In addition, the program uses the
built-in {\tt KNOWN} that, when used on unknowns, checks whether the
domain of the argument is a singleton.

\small
\begin{verbatim}
TYPE Colour = (blue, green, red, yellow);
     Info =  RECORD 
       co: Colour; 
       No: CONSTRAINED INTEGER;
     END;
     Board = ARRAY [1..M],[1..N] OF Info;

PROCEDURE Region(VAR X: Board; VAR number: INTEGER); 
  VAR i, j, k: INTEGER;
BEGIN
  FOR i := 1 TO M DO 
    FOR j := 1 TO N DO 
      IF i < M AND X[i,j].co = X[i+1,j].co 
      THEN X[i,j].No = X[i+1,j].No 
      END;
      IF j < N AND X[i,j].co = X[i,j+1].co 
      THEN X[i,j].No = X[i,j+1].No 
      END
    END
  END;
  k := 0;
  FOR i := 1 TO M DO 
    FOR j := 1 TO N DO 
      IF NOT KNOWN(X[i,j].No) 
      THEN k := k+1; X[i,j].No = k
      END
    END
  END;
  number = k
END Region;
\end{verbatim} 
\normalsize

Note that for any {\tt i} in {\tt [1..M]} and {\tt j} in {\tt [1..N]},
the record component {\tt X[i,j].No} is of a constrained type.  Here
the first double {\tt FOR} statement generates the constraints while
the second double {\tt FOR} statement solves them by assigning to the
pixels that belong to the same region the same number. 

Due to the call by variable mechanism, the actual parameter
corresponding the formal one, {\tt X}, is modified by the procedure.
In particular, the second component, {\tt No}, of each array element is
instantiated after the procedure call.

Next, we explain the call by value mechanism in presence of
unknowns.  An unknown passed as an actual value parameter is treated
as a customary variable: it is evaluated and its value is assigned to
a local variable associated with the formal parameter.  If the value
of this unknown is at this moment undetermined, this evaluation yields
a run-time error.  This evaluation process also applies if a field or
an element of a compound actual value parameter is an unknown.


\section{Language Extensions}\label{sec:extensions}

In this section we discuss some built-in procedures of the proposed
language
that make it easier for the user to program with constraints.  In
particular,
in Subsection~\ref{subsec:constraint-built-ins} we discuss built-ins for
stating constraints, and in Subsection~\ref{subsec:indomain-built-in} we
present built-ins for assigning values to unknowns.

\subsection{Built-ins for Expressing
Constraints}\label{subsec:constraint-built-ins}

The practice of constraint programming requires inclusion in the
programming language of a certain number of language built-ins that
facilitate constraint formulation.

For example, if we wish to state that the unknowns of the array {\tt
  X} must have pairwise different values, we write

\begin{small}
\begin{verbatim}
   ALL_DIFFERENT(X);
\end{verbatim}
\end{small}

This call results in a constraint which is equivalent to the set of
all the corresponding constraints of the form {\tt X[i] <> X[j]}, for
{\tt i} $\in${\tt [1..N-1]} and {\tt j} $\in${\tt [i+1..N]}.
\footnote{In some systems, such a constraint is kept in its original
  form in order to exploit constraint propagation techniques that deal
  specifically with constraints of this kind, see \cite{Reg94}.}

Similarly, if we wish to state that at most {\tt k} among the unknowns
belonging to the array {\tt X} can have the value {\tt v}, we write

\begin{small}
\begin{verbatim}
   AT_MOST(k,X,v);
\end{verbatim}
\end{small}

This sort of built-ins on arrays are present in other imperative
constraint languages.  We do not list all of them here, but we
envision their presence in the language.

Such built-ins on arrays are the counterparts in imperative languages
of the corresponding built-ins on lists provided by constraint logic
programming systems such as CHIP.
These languages also support symbolic manipulation of terms which makes
it
easy to generate arithmetic constraints. The traditional imperative
programming languages lack this power and exclusive reliance on arrays
can lead to artificial and inefficient solutions.

For example, suppose we are given an ${n\times n}$ matrix $A$ of
integer unknowns and we wish to state the constraint that the sum of
the elements of the main diagonal must be equal to a given value
$b$.  A customary solution would involve resorting to an auxiliary array
of unknowns in the following way:

\begin{small}
\begin{verbatim}
VAR A: ARRAY [1..N], [1..N] OF CONSTRAINED INTEGER;
    V: ARRAY [1..N] OF CONSTRAINED INTEGER;
    b: INTEGER;

   V[1] = A[1,1];
   FOR i := 2 to N DO
      V[i] = A[i,i] + V[i-1];
   END;
   V[N] = b;
\end{verbatim}
\end{small}

This solution, which one would write for example in ILOG Solver, has the
obvious drawback of creating {\tt N} new unknowns for stating one single
constraint.

Therefore we propose the use of lists of unknowns (as done for example
in the ICON programming language of \cite{GG83} for the case of
variables), identified by the keyword {\tt LIST}, upon which
constraints of various forms can be stated by means of built-ins.  The
above program fragment would then be replaced by

\begin{small}
\begin{verbatim}
VAR A: ARRAY [1..N], [1..N] OF CONSTRAINED INTEGER;
    L: LIST OF CONSTRAINED INTEGER;
    b: INTEGER;

   Empty(L);
   FOR i := 1 to N DO
     Insert(L, A[i,i])
   END;
   Sum(L,'=',b);
\end{verbatim}
\end{small}

\noindent where {\tt Sum} is a built-in with the expected meaning of constraining the sum of the unknowns in {\tt L} to be equal to $b$. Once
the constraint {\tt Sum(L,'=',b)} has been added to the store, the
variable {\tt L} can be used again for a different purpose.  Note that
in this solution no additional unknowns are created.  In order to
obtain a similar behaviour in ILOG Solver one needs either to add a
similar built-in to it or to make explicit use of pointers to objects
representing unknowns.

Consider now again the Frequency Assignment problem. We discuss here
the formulation of an additional constraint for this problem which
requires the use of lists. Suppose that we wish to state that in a
particular region (i.e., a set of cells) a given frequency is used no
more than a given number of times.

This type of constraint is useful in real cases. In fact, in some
situations even though the pairwise interference among cells is below a
given threshold and no separation is required, the simultaneous use of
a given frequency in many cells can create a interference phenomenon,
called {\em cumulative interference}.

The following procedure states the constraints for preventing
cumulative interference in region {\tt R} (where the type {\tt Region}
is an array of booleans representing a subset of the set of cells).
Here {\tt max} is the maximum number of cells in the region that can
use the same frequency.

\begin{small}
\begin{verbatim}
PROCEDURE RegionConstraint(R: Region; max: INTEGER; VAR X: Assignment); 
VAR i, k: INTEGER;
    L: LIST OF CONSTRAINED [1..M];

BEGIN
  FOR k := 1 TO M DO
    Empty(L);
    FOR i := 1 TO N DO
      IF R[i] THEN Insert(L,X[i]) END
    END;
    AT_MOST(max,L,k)
  END
END RegionConstraint;
\end{verbatim}
\end{small}

\subsection{Built-ins for Assigning
Values}\label{subsec:indomain-built-in}

In order to search for a solution of a set of constraints, values must
be assigned to unknowns. We define the built-in procedure {\tt
  INDOMAIN} which gets an unknown of a finite type (so {\tt BOOLEAN},
enumeration or a subrange type) as a parameter, and assigns to it {\em
  one\/} among the elements of its domain.  The procedure also creates
a choice point and all other elements of the domain are successively
assigned to
the unknown upon backtracking.

The choice of the value to assign to the unknown is taken by the
system depending on the current state of the store, based on
predefined {\em value selection\/} strategies. We do not discuss the
issue of which are the best value selection strategies. We only assume
that all consistent values are eventually generated, and that the
choice point is erased after the last value has been generated.

The procedure {\tt INDOMAIN} can be also used on arrays and on lists.
For example, the call {\tt INDOMAIN(A)}, where {\tt A} is a matrix of
integer unknowns, generates (upon backtracking) all possible
assignments for all elements of {\tt A}.

The order of instantiation of the elements of {\tt A} is taken care of
by the store, which applies built-in strategies to optimize the
retrieval of the first instantiation of the unknowns. As in the case
of value selection, we do not discuss here the issue of the {\em
variable ordering}.

\section{Related Work}\label{sec:related-work}

We concentrate here on the related work involving addition of
constraints to imperative languages.  For an overview of related work
pertaining to the \almazero{} language we refer the reader to
\cite{ABPS98a}.

As already mentioned in the introduction, the most successful imperative
constraint language is the C++ library ILOG Solver \cite{ILOG98}.
The main difference between our proposal and ILOG Solver is that the
latter is
based on the conventional imperative language C++ and consequently it
does not
support automatic backtracking. Therefore the interaction with the store
cannot
be based on failures issued by the store constraint solvers while
evaluating
the statements. In ILOG Solver such an interaction is always explicit,
whereas
in our proposal we aim at making it transparent to the user.


We are aware of two other language proposals in which constraints are
integrated into an imperative language --- the commercial language
CHARME of \cite{OPLOBEDU89} and 2LP of \cite{MT95b}.  In
each language some of the issues here discussed have been addressed,
but not all of them. 

More specifically, in CHARME unknowns (called logical variables) and
linear constraints on them are allowed. The language supports use of
Prolog-like terms, arrays and sequences of logical variables and a
number of features (like demons and the {\tt element} primitive, an
equivalent of {\tt INDOMAIN}) adopted from the CHIP language.  Also,
it provides a nondeterministic {\tt or} statement and iterations over
finite domains, arrays and sequences of logical variables.

The C like syntax creates an impression that CHARME supports
imperative programming.  However, from the paper it is not clear
whether it is actually the case.  If it is, then it is not clear how
the logical variables, constraints and nondeterministic statements
interact with the usual features of the underlying imperative
language.  In particular, the use of logical variables outside of
constraints, the impact of backtracking on the assignment statements
and the status of choice points created within procedure bodies is not
explained (probably due to space limitations).  CHARME does provide
bidirectional connection with C.

2LP was designed for linear programming applications.
In 2LP unknowns (called {\it continuous\/} variables) are global. They
vary over the real interval $[0, + \infty)$ and can be either simple
ones or arrays.  The only way these variables can be modified is by
imposing linear constraints on them. Constraints can also appear in
conditions. This leads to a conditional way of adding them to the store.

Whenever a constraint is added to the store, its feasibility w.r.t.
the old constraints is tested by means of an internal simplex-based
algorithm. This algorithm maintains the current feasible region,
which is a polyhedron, together with a {\em witness point\/} which is
a distinguished vertex.

The continuous variables can appear outside of the constraints as
arguments of any procedure whose signature has a continuous variable,
and
as arguments to some predeclared functions like {\tt wp} that returns
the value of a witness point.  In the latter case when a continuous
variable is passed as a parameter, the witness point value is used.

2LP provides the nondeterministic statements analogous to the {\tt
  ORELSE} and {\tt SOME} statements of \almazero{} and a limited form
for the {\tt FORALL} statement. Automatic backtracking over assignment
and combination of continuous and customary variables in compound
variables is not supported.

%
%

\section{Conclusions and Future Work}\label{sec:conclusions}

In this paper we discussed the programming language \almazero{} that
integrates the imperative and logic programming paradigm and
illustrated the resulting programming style by a number of examples.
\almazero{} is based on first-order logic in the sense that it provides a
computational interpretation for the standard connectives, so
negation, disjunction and conjunction, and for various forms of 
quantification.
In fact, many first-order formulas and their extensions by bounded
quantifiers, sorts (i.e., types), and arrays, can be interpreted and
executed as \almazero{} programs.  The precise logical nature of this
computational interpretation of first-order logic was worked out in
\cite{AB99}.

Then we discussed a proposal how to integrate constraint programming
features into the language.  In this regard we believe that the use of
an underlying language based on first-order logic, such as
\almazero{}, rather than a conventional imperative language, makes the
integration of constraints more natural and conceptually simpler.

We analyzed here a number of issues related to the proposed integration,
such as the use of constrained types and the unknowns, interaction
between the program and the constraint store, and the parameter
passing mechanisms.  Finally, we presented some examples that
illustrate the resulting style of programming.

In our future work we plan to extend the work carried out in
\cite{ABPS98a} to the language proposal here outlined.
More specifically, we envisage to
\begin{itemize}
\item extend the executable, operational semantics based on
  the ASF+SDF Meta-Environment of \cite{Kli93};
\item extend both the \almazero{} compiler and its underlying abstract
  machine \aaa{};
\item implement a set of constraint solvers or provide an interface
  between the language and existing constraint solvers.
\end{itemize}

The first item can be dealt with by adding to the executable semantics
of \almazero{} given in \cite{ABPS98a} a few rules that
formalize the interaction between the program and the store stipulated
in Subsection~\ref{subsec:store-vs-program}.  These rules are
parameterized by the constraint solvers attached to the store.

Regarding the last item, we plan to develop a simple solver for
constraints over finite domains to be used for prototyping and testing
purposes. We also plan to exploit more powerful external solvers
already available for subsequent releases of the system.

As already mentioned in Section~\ref{subsec:store-vs-program}, we do
not allow so-called {\em ask} operations in the store.  This is a
deliberate design decision which allows us to keep the language 
design simple and the underlying execution model easy to implement.

Nevertheless, in future versions of the language, we plan to
investigate the possibility of equipping the store with an {\em
  entailment} procedure.  This procedure should check whether an
evaluated form of a constraint is logically implied (or {\em
  entailed}) by the store.  Upon encounter of an {\tt ask} constraint,
the entailment procedure would check whether the evaluated form is
entailed by the store. If it is the case, the constraint evaluates to
{\tt TRUE}.  Otherwise the constraint evaluates to {\tt FALSE}.  We
would require that the entailment procedure returns correct results
but would not assume that it is complete.

We did not deal here with some of the issues related to the design of
the language. Specifically, we omitted discussion of

\begin{itemize}
\item a full set of built-ins, in particular the ones appropriate for
  constraint optimization,
\item primitives for selecting variable and value selection strategies,
\item the language support for the dynamic creation of unknowns.
\end{itemize}

These can be taken care of in a systematic way and lead to a complete
and rigorous definition of an imperative constraint programming
language.

\section*{Acknowledgements}

We would like to thank Jan Holleman, Eric Monfroy and Vincent
Partington for useful discussions on the subject of this paper.
Helpful comments by Tony Hoare and other two, anonymous, referees
allowed us to improve the presentation.

\bibliographystyle{plain}

\bibliography{/ufs/apt/book-ao-2nd/apt,/ufs/apt/esprit/esprit,/ufs/apt/bib/clp2,/ufs/apt/bib/clp1,/ufs/apt/book-lp/man1,/ufs/apt/book-lp/man2,/ufs/apt/book-lp/man3,/ufs/apt/book-lp/ref1,/ufs/apt/book-lp/ref2,/ufs/apt/bib/a0}

\end{document}